# Who are we now?

In 2014 the Royal Astronomical Society carried out a survey of its membership, finding that we are both more and less diverse than UK society as a whole. **Robert Massey** summarizes the findings and what they mean for the Society in future.

In recent years, the demographics of our Fellows – especially their diversity – has been an area of concern for the Royal Astronomical Society, in common with other learned societies. Officers, councillors and staff, as well as the wider fellowship, have extended existing efforts to tackle the under-representation of women in the Society and in astronomy and geophysics; we are also working to better represent people from a range of ethnicity, disability and socioeconomic backgrounds.

Three significant changes have taken place to reflect this new emphasis, namely the creation of the Committee on Diversity in Astronomy and Geophysics (CDAG), the recruitment of Sheila Kanani as our first dedicated Education, Outreach and Diversity Officer, and the signing by the RAS of the Science Council Charter on Diversity, Equality and Inclusion in October 2014.

The charter commits the Society to take action; the first was the appointment of Prof. Jim Wild as Diversity Champion, an advocate for this work on RAS Council. But the charter also requires us to gather data on our membership and on the wider astronomy and geophysics communities, to identify areas where we fall short of matching the demographic composition of wider society, look at barriers to change and develop a programme to address them.

In 2010 the Society commissioned a general demographic survey of the UK astronomy and geophysics research community which, while focused on research interests, also asked respondents about characteristics such as sex and ethnicity for the first time. This dataset demonstrated that the proportion of women in astronomy and geophysics is rising slowly at all levels, from a low base, but that people from black and minority ethnic backgrounds are greatly under-represented compared with the population as a whole.

We carried out a further survey of our membership in the summer of 2014, with guidance from CDAG, Council and other learned societies, to find out more about the diversity of our Fellows and help shape future public engagement. Questions in the survey covered country of residence, age, sex, ethnicity, disability, religion and sexual orientation. Respondents were also invited to comment. The questions were chosen after consultation with staff at the Institute of Physics and the Equality Challenge Unit (ECU), the body which works to further and support equality and diversity for staff and students in higher education across the UK. ECU categories were used where possible, though some subcategories (e.g. denominations in Christianity) were merged for simplification. The questions in most cases also allow direct comparison with UK census data from 2011 and with Office of National Statistics (ONS) data on sexual orientation from 2006.

To carry out the survey, the Society bought Snap Surveys software, which ensured that all electronic responses remained anonymous (to Society staff and anyone else); access to the questionnaire was via a web link. A draft questionnaire assembled by Fern Storey, then Membership Secretary, was tested with staff in-house and with the chair of CDAG, all of whom assisted in framing the questions. The survey went live on 16 July 2014 and was promoted to 3486 of our Fellows via email; a paper copy was sent to a further 457 by post. The survey closed on 2 September 2014, by which time 1444 Fellows had responded (including 68 by post), a rate of 36.6%. The Membership Secretary then worked on the data. Results were presented to CDAG and the Membership Committee.

### Residence
In line with information in the RAS membership database, 1013 respondents (70%) lived in the UK. Among those resident elsewhere, the highest proportion were in the United States (11%). 129 respondents (9%) were resident in EU countries and UK crown dependencies such as Jersey and the Isle of Man. There was only 1 respondent from China, 10 from India, 1 from Taiwan, 9 from South Africa and 3 from Japan, despite the growing research base in the first three of these countries and the established global strength of the other two.

### Age
The age of respondents varied from 21 to over 100; the mean age of UK respondents was 54 and the median age 55. Members outside the UK are slightly older, with a mean age of 58 years and a median of 60. The distribution was skewed towards older age groups within and outside the UK; there is an increase in the number of respondents as the age rises. The Society also has many retired Fellows, with 296 respondents aged 70 and over.

Within that overall pattern, there is a noticeable dip from the 25–29 years band (116 respondents) down to 35–39 years (85 respondents), before the upward trend continues. This may reflect the age when many Fellows who have completed one or two postdoctoral positions move into employment outside academic science and resign their membership of the Society. It also corresponds to the point when early-career Fellows lose their entitlement to a reduced annual subscription, perhaps making continued membership less appealing.

> "The 2010 survey showed the proportion of women is rising slowly at all levels"

For comparison, our 2010–2011 survey identified a median age of 40–44 years for UK staff working in astronomy and solar system science (there were too few respondents in solid-Earth geophysics for statistical significance). This is very different to the age profile of the Society, but does not include retired researchers.

### Sex
80% of respondents stated that they were male and 19% female, with 0.3% choosing "other" and 0.7% indicating "I do not wish to say". Among UK Fellows who responded, 21% were women and 77.5% were men. The RAS membership database indicates that 16% of Fellows are female, indicating that women were slightly more likely than men to respond to the survey.

There is a significant difference in the proportion of female members in different age ranges. Among UK members in their 20s and 30s, women make up 38–44% of those responding, a relatively healthy level and stronger than across academic physics in general (see e.g. the Institute of Physics report *Academic Physics Staff in UK Higher Education Institutions*). Beyond the age of 40 there is a steady decline, with the proportion of women dropping to 20% for the

## 1 Age

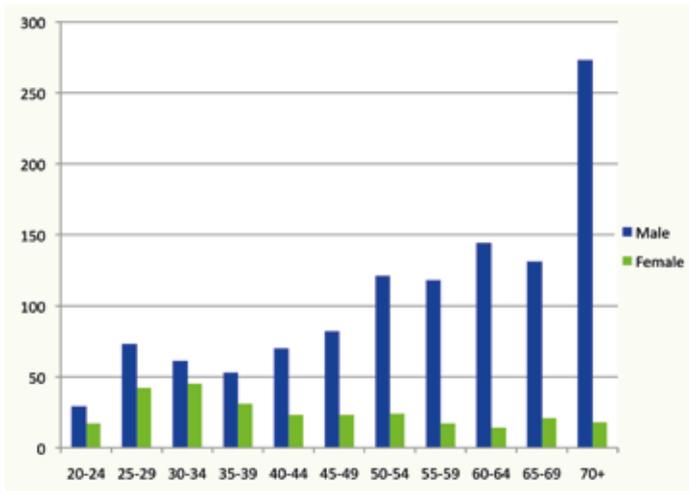

Age distribution among all Fellows who responded, divided for male and female. Ten respondents chose not to identify as male or female and four identified as other; these respondents have not been included in this plot.

## 2 Disability

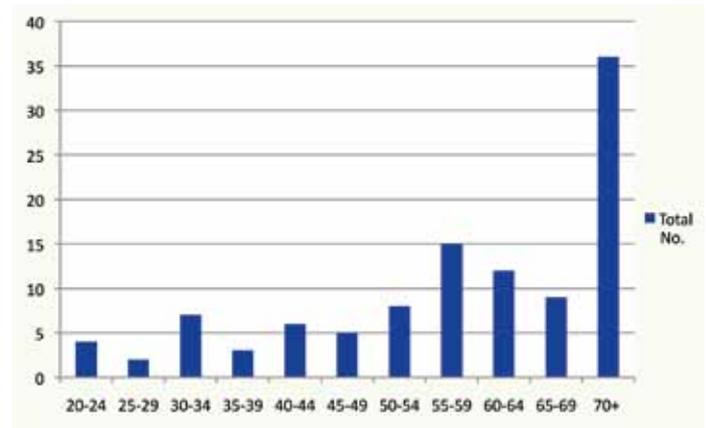

Numbers of UK respondents with a disability. 107 Fellows responded, comprising 10.6% of the sample.

45–49 year old cohort. The absolute number and proportion of women members continues to decline with age; female respondents make up just 6% of those over 70.

The 2010–2011 survey found lower proportions of women in universities at all levels: around 34% of research students in the UK were women; women made up 34% of academic staff aged 25–29 and around 25% of those aged 30–39; the proportion declined to around 10% after the age of 45.

Academia is, however, only one area where Fellows work; many are employed in science centres and museums and in careers outside astronomy and geophysics. Nonetheless, given the pool of candidates for Fellowship, most of whom at least begin careers in research, the Society appears to be significantly better at recruiting women than men, under the age of 40. This is, however, not yet reflected in elections to RAS Council, where women typically only make up between a fifth and a quarter of nominated candidates.

**Ethnic origin**
In the 2010–2011 survey, the Society asked UK researchers to state their ethnicity for the first time. From these data it became clear that there are relatively few PhD students and employees in our fields from minority ethnic backgrounds, far fewer than in the UK population as a whole.

This is reflected in the Fellowship of the RAS in 2014. Of the 1013 respondents from the UK, 7.2% stated that they were from an ethnic minority (this includes 1.3% who identified as "other white" such as European nationals). In the most recent UK census (2011), 19.5% of the population were in these categories. Within the Society, some groups are exceptionally under-represented. For example, just 4 respondents (0.6% of the UK sample) gave their ethnicity as African, Caribbean or Other Black, compared with 2.9% in the population as a whole. Respondents who gave their ethnicity as Indian made up 1.8% of the sample, compared with 2.3% of the UK population. For Fellows resident outside of the UK, 87% of those who answered this question were white, reflecting the dominance of Europe and North America in our overseas membership.

**Disability**
This survey marked the first time that the Society has sought data on disability and impairment. As well as being in keeping with our general approach to diversity and inclusion, these data allow us to consider better provision for the needs of our membership. ECU categories informed the questions, which started by asking respondents whether they had a disability, health condition or learning difference. 10.6% of UK members replied "yes" and were then asked to identify one or more areas. Unsurprisingly the incidence of disability and impairment rises with age, with UK respondents indicating they had one or more condition rising from 4% of those aged 20–29 to 20.2% of those 70 and over.

In the UK census for 2011, 18.2% of respondents reported a condition which led to some degree of limitation on their day-to-day activities. RAS UK members thus appear to be significantly less likely to have a long-term limiting health condition or disability than the wider population. Of Fellows resident outside the UK, only 5.3% reported a similar condition.

Overall, 12 people (0.8% of the total respondents) indicated they had a specific learning difficulty, and 6 people (0.4%) a social or communication impairment such as Asperger's syndrome. 32 respondents (2.2%) reported a long-standing illness, 15 (1%) a mental health condition and 22 (1.5%) a physical impairment or mobility issue. 16 respondents (1.1%, all male) reported deafness and 4 (0.3%) blindness.

**Sexual orientation**
This question was included after discussions with the Institute of Physics, which had carried out a survey that did not ask for this information. IOP members had then asked why these data were not collected, despite being a "protected characteristic" in UK law. Both the IOP and RAS have members in countries where disclosure of sexual orientation brings a risk of prosecution and harassment; although the survey was anonymous, residents in those states who are not heterosexual would have good reason not to answer this question. Interestingly, this does not appear to have had a significant impact on this survey: 6% of respondents outside the UK and 7% in the UK refused to answer, although this may simply reflect the dominance of North American and European residents in our overseas membership.

Among UK respondents, 84% identified as heterosexual, 3% as bisexual, 4% as gay men, 0.2% as lesbian and 1% as "other", while 7% did not wish to say. Data on sexual orientation have not been collected in the UK census, but the ONS has developed and tested appropriate questions since 2008. These data indicate that, across the UK, 1.5% of men identify as gay and 0.7% of women identify as lesbian, with 0.3% of men and 0.5% of women identifying as bisexual. A further 4.7% did not answer.

These data thus suggest that the Society has more gay men than the population as a whole. On the other hand, the number of

## 3 Sexual identity

| sex (UK residents) | no. | % |
|---|---|---|
| female | 216 | 21.3 |
| male | 785 | 77.5 |
| other | 3 | 0.3 |
| I do not wish to say | 9 | 0.9 |
| sex (non-UK residents) | no. | % |
| female | 59 | 13.7 |
| male | 370 | 85.8 |
| other | 1 | 0.2 |
| I do not wish to say | 1 | 0.2 |

women identifying as lesbian is very small, significantly lower than the proportion in the ONS survey data. A higher proportion of RAS Fellows in the UK identify as bisexual compared to the wider population.

Outside the UK, 6% of respondents gave their sexual orientation as bisexual, 1% as gay men, 0.2% as lesbian and 85% as heterosexual.

### Religion

Religious belief is a standard census question in the UK. There are differences between the countries with, for example, Scotland having a significantly higher proportion of people with no religion (36.7%) than England and Wales (25.1%) and Northern Ireland (13.9%).

Within the UK, 59.9% of responding RAS Fellows indicated they hold no religious belief. Of the remainder in the UK, 31.9% described themselves as Christian, 0.3% Muslim and 0.1% as Sikh. The latter three faiths have a much lower occurrence than in the population at large. With the caveat that this is a very small sample, Buddhism (0.8%) and Judaism (0.7%) are slightly over-represented in the fellowship, when compared with the census data.

RAS Fellows outside the UK are slightly more religious. 53.1% of respondents describe themselves as holding no religious belief. 36.7% are Christian, 2.3% Muslim, 1.9% Hindu, 0.7% Jewish, 0.7% Buddhist and 0.2% Sikh. Overall, men were a little more likely (59.0%) than women (54.9%) to hold no religious belief, but there were no significant differences in the proportions of men and women adhering to specific faiths.

### Comments from Fellows

This was the first time a survey had taken place of protected characteristics of the RAS Fellowship. As noted earlier, the survey attracted a good response rate. We invited respondents to comment: 236 did so, anonymously. While some questioned the value of a diversity survey, many others favoured this new approach and suggested examination of further categories such as

## 4 Ethnicity

| UK | | RAS |
|---|---|---|
| 82.3 | White (British) | 91.3 |
| 0.8 | Irish | 0 |
| 0.1 | Gypsy or Irish Traveller | 0 |
| 3.9 | Other White | 1.3 |
| 0.7 | White and Black Caribbean | 0.2 |
| 0.3 | White and Asian | 0.7 |
| 0.5 | White and Black African | 0 |
| 0.5 | Other Mixed background | 1.1 |
| 2.3 | Indian | 1.8 |
| 1.9 | Pakistani | 0 |
| 0.7 | Bangladeshi | 0.1 |
| 0.7 | Chinese | 0.2 |
| 1.4 | Other Asian | 0.5 |
| 1.6 | African | 0.2 |
| 0.9 | Caribbean | 0.1 |
| 0.4 | Other Black | 0.1 |
| 0.4 | Arab | 0.2 |
| 0.5 | Any other ethnic group | 0.8 |

Percentages identifying with standard ethnic classifications for the UK as a whole (based on 2011 census returns) and for this RAS survey.

social background. Constructive suggestions for the categories and ordering of questions will be taken into account in the design of the next survey of this kind.

### Conclusions and recommendations

The survey has demonstrated that the RAS is a diverse organization in that it represents people of many different backgrounds, even if this is not always apparent at its meetings. But the Society does not reflect the ethnicity and sex of the wider population – something it has in common with academic astronomy and geophysics in the UK. The RAS also has a significantly older profile than the research community, with a dip in the number of members between the ages of 30 and 40.

These data pose a series of challenges for the Society that will occupy CDAG and Council for some time to come. We would welcome suggestions for ways forward from Fellows, many of whom are active in initiatives such as Project JUNO. Areas already under discussion include:
● Strengthening our nascent work on engaging people from black and minority ethnic backgrounds, as these groups are very poorly represented in both academic research in astronomy and geophysics and in the RAS. This is a high priority area for both CDAG and for our Education and Outreach Committee.
● Looking at measures to encourage more of the relatively high fraction of younger women in the Society to take up committee

## 5 Religion

| UK % | | UK RAS % |
|---|---|---|
| 0.4 | Buddhist | 0.9 |
| 59.5 | Christian | 31.9 |
| 1.3 | Hindu | 0.7 |
| 0.4 | Jewish | 0.8 |
| 4.4 | Muslim | 0.3 |
| 0.7 | Sikh | 0.1 |
| 0.4 | other | |
| 7 | not stated | 5.4 |
| 25.8 | no religion | 59.9 |

posts or seek election for Council.
● Being mindful of the religious diversity of the Society and also considering how to use this diversity to good effect.
● Working to establish why people with disabilities are under-represented in the Society, and explore how this might be mitigated, for example in access arrangements for meetings and conferences.
● Moving to implement a Diversity Champions project across the UK, with RAS funding for researchers to carry out public engagement work in areas of socio-economic deprivation and ethnic diversity.
● Examining the potential for recruitment from outside Europe and North America and considering what might encourage people in these regions to join the Society.
● Investigating the feasibility of an exit survey to find out why Fellows leave.
● Sharing and drawing on good practice in diversity policy and operational management with other organizations, for example, offering staff training in areas such as unconscious bias and developing an equal opportunities policy.
● Exploring the greater use of remote access to meetings to facilitate involvement of parents and others with caring responsibilities.
● Defraying the cost of additional childcare for RAS council members and others attending committee meetings or on general Society business.
● Providing crèche facilities at NAM and other large meetings where the Society has significant involvement. ●

**AUTHOR**
**Robert Massey** is Deputy Executive Secretary of the RAS.